\documentclass[twocolumn,aps,floats,superscriptaddress,showpacs]{revtex4}
\usepackage{amsmath}

\usepackage{epsfig}
\usepackage{graphicx}
\usepackage{dcolumn}
\usepackage{bm}

\def\gapp{\lower.35em\hbox{$\stackrel{\textstyle>}{\sim}$}}
\def\lapp{\lower.35em\hbox{$\stackrel{\textstyle<}{\sim}$}}

\begin{document}

\title{Electronic properties of curved graphene sheets}
\author{Alberto Cortijo}
\affiliation{Unidad Asociada CSIC-UC3M,
Instituto de Ciencia de Materiales de Madrid,\\
CSIC, Cantoblanco, E-28049 Madrid, Spain.}

\author{ Mar\'{\i}a A. H. Vozmediano}
\affiliation{Unidad Asociada CSIC-UC3M,
Universidad Carlos III de Madrid, E-28911
Legan\'es, Madrid, Spain. }

\date{\today}
\begin{abstract}
A model is proposed to study the electronic structure of slightly curved graphene
sheets with an arbitrary number of pentagon-heptagon
pairs and Stone-Wales defects based on
a cosmological analogy. The disorder induced by curvature produces
characteristic patterns in
the local density of states that can be observed in scanning tunnel and
transmission electron microscopy.

\end{abstract}
%

\maketitle
The recent synthesis of single or few layers of graphite\cite{Netal05,Zetal05}
allows to test the singular transport properties predicted in early
theoretical studies\cite{S84,Hal88,GGV96,Khv01} and experiments\cite{Ketal03}.
The discovery of a
substantial field effect \cite{Netal04} and of ferromagnetic
behavior\cite{Eetal03} allows to envisage graphene as a reasonable
replacement of nanotubes in electronic applications.
Disorder plays a very important role in the electronic
properties of low dimensional materials.

Substitution of an hexagon by an n-sided ring in the hexagonal lattice
without affecting the threefold coordination of the carbon atoms leads to the warping
of the graphene sheet. Rings with $n>6$ $(n<6)$, induce locally positive (negative)
curvature.
Inclusion of an equal number of pentagons and heptagonal rings
would keep the flatness of the sheet at large scales and produce
a flat structure with curved portions that would be structurally stable
and have distinct electronic properties.
This defects give rise to long range modifications
in the electronic wave function  that affect the electronics
in a way different
from that produced by  vacancies or other impurities modelled by local potentials.
Pentagon-heptagon pairs and
Stone-Wales defects made of two adjacent heptagons and two pentagons
form naturally in experiments of ion bombarded nanotubes as a mechanism
to reduce the dangling bonds in large vacancies\cite{Aetal98} and
have been observed to form in situ
in single graphene layers by high-resolution transmission electron
microscopy (TEM)\cite{Hetal04}.

We propose a method, based on a cosmological analogy, to compute the
electronic structure and transport properties
of curved graphene sheets consisting of
an arbitrary number of topological defects
located at fixed positions of the lattice. We see that, unlike
vacancies and voids, the combination of positive and negative
curvature gives rise to characteristic
inhomogeneous patterns in the density of states that
affect the transport properties of the layers and can be observed
in scanning tunnel (STM) and
electron transmission spectroscopy (ETS). The present analysis can help
in the experimental  characterization of graphene samples
and in the correct analysis of STM images.
The results obtained can be related to
recent Electrostatic Force Microscopy (EFM) measurements
that indicate large potential differences
between micrometer large regions on the surface of highly oriented graphite\cite{GE06}
and to the suppression of magnetorresistance in
single layered graphene reported in \cite{Metal06}.
We apply the formalism to present the corrections to the local
density of states induced by pentagon-heptagon pairs and Stone-Wales
defects in the average planar graphene sheet.

{\it Description of the model.}
The conduction band of graphene is well described by a tight
binding model which includes the $\pi$ orbitals
which are perpendicular to the plane at each
C atom\cite{W47,SW58}. This model describes a semimetal, with zero
density of states at the Fermi energy, and where the
Fermi surface is reduced to two inequivalent K-points
located at the corners of the hexagonal
Brillouin Zone.
The low-energy excitations with momenta
in the vicinity of any of the Fermi points $K_{\pm}$
have a linear dispersion
and can be described by a continuous model which reduces to the
Dirac equation in two dimensions\cite{GGV92,GGV93,GGV94}.
In the absence of interactions or disorder mixing  the two
Fermi points the electronic properties of the system are well described
by the effective low-energy Hamiltonian:
\begin{align}
{\cal H}_{0\pm}= i\hbar v_{\rm F}
( \sigma_x \partial_x \pm  \sigma_y \partial_y)
\;,\label{hamil}
\end{align}
where $\sigma_{x,y}$ are the Pauli matrices,
$v_{\rm F} = (3 t a )/2 $, and $a=  1.4 \AA$ is the
distance between nearest carbon atoms. The
components of the two-dimensional spinor:
$\bar{\Psi_i}( {\bf r} )= (
\varphi_A ( {\bf r} ),\varphi_B ( {\bf r }))_i$
correspond to the amplitude of the wave function in each of the two
sublattices (A and B) which build up the honeycomb structure.
Pentagons and heptagons can be viewed as disclinations
in the graphene lattice, and, when circling one such defect, the
two sublattices in the honeycomb structure as well as the
two Fermi points are exchanged.
The scheme to incorporate this change in a continuous
description was discussed in refs. \cite{GGV92,GGV93} and \cite{GGV01}.
The process can be described by means
of a non Abelian gauge field, which rotates the spinors in the
flavor space of the Fermi points. The two spinors associated to
each Fermi point can be  combined
into  a four component Dirac spinor $\bar{\Psi}( {\bf r} )= (
\Psi_+ ( {\bf r} ),\Psi_- ( {\bf r }))$ which in
%
the presence of a single disclination is an eigenstate  of the
Hamiltonian\cite{CV06}
\begin{eqnarray}
H=-i v_{_{F}}{\vec \gamma}.{\vec\partial}+g\gamma^{q}
{\vec \gamma}.{\vec A}({\bf r}). \label{hamgauge}
\end{eqnarray}
$v_{_{F}}$ is the Fermi velocity,
$ \gamma^{i}$ are
$4\times4$ matrices constructed from the Pauli matrices,
$\gamma^q=\frac{\sigma_3}{2}\otimes I$,
the latin indices run over the two spatial dimensions and
g is a coupling parameter related to the deficit angle of the  defect.

We  use time-independent
perturbation theory to calculate corrections to the self-energy
in the weak coupling regime of the parameter $\widehat{g}\equiv
\frac{\Phi}{2\pi L^2},\label{singlecoupling}$ where the constant
$\Phi$  is the strength of the vortex: $\Phi=\oint\vec{A}d\vec{r}
$  related to the opening angle of the defect by $\Phi=(1-b)$. L
is the dimension of the sample.

In a disordered system, in general, because of the presence of an
external potential or impurities, the space is inhomogeneous. The
Green's function doesn't depend on the difference ($\bf{r}-r'$)
and $\bf{k}$ is no longer a good quantum number. If the external
potential or the effect of the impurities are time-independent we
have elastic scattering and the states \textbf{k} and \textbf{k}'
have the same energy. We want to calculate the total density of
states $\rho(\omega)$ of the system perturbed by the defect via
the vector potential given in eq. (2). This density is the
imaginary part of the Green's function integrated over all
positions, in the limit $\textbf{r}'\rightarrow \textbf{r}$:
\begin{equation}
\rho(\omega)=\int Im
G(\omega,\textbf{r},\textbf{r})d\textbf{r}.\nonumber
\end{equation}
In terms of the Green's function in momentum representation,
$\rho(\omega)$ can be written as:
\begin{equation}
\rho(\omega)=Im\int\int \frac{d\textbf{k}}{(2\pi)^{2}} \int
\frac{d\textbf{k}'}{(2\pi)^{2}}
e^{i\textbf{kr}}e^{-i\textbf{k}'\textbf{r}}G(\omega,\textbf{k},\textbf{k}')d\textbf{r}.
\nonumber
\end{equation}
The integration over $\textbf{r}$  gives delta function
$4\pi^{2}\delta(\textbf{k}-\textbf{k}')$, and $\rho(\omega)$ then
reads:
\begin{equation}
\rho(\omega)=\int \frac{d\textbf{k}}{(2\pi)^{2}}Im
G(\omega,\textbf{k},\textbf{k}).\nonumber
\end{equation}
%

{\it Generalization to negative curvature and an arbitrary number
of defects.} An alternative approach to the gauge theory of
defects  is to include the local curvature induced by an
n-membered ring by coupling the Dirac equation to a curved space.
This approach was introduced in \cite{GGV92,GGV93} to study the
electronic spectrum of closed fullerenes and has recently been
used for fullerenes  of  different shapes in \cite{PPO06}. The
holonomy discused in the gauge approach is included in the
formalism by means of the spin connection\cite{FMC06}. In most of
the previous works in this context, the graphene sheet was wrapped
on a geometrical surface that can be easily parametrized. The
situation is more complicated if we try to describe the samples
that are being obtained in the laboratory that are flat sheets
with local corrugation\cite{Metal06}.

In this context one can see that the substitution of an hexagon by
a polygon of $n<6$ sides gives rise to a conical singularity with
deficit angle $(2\pi/6)(6-n)$. This kind of singularities have
been studied in cosmology as they are produced by  cosmic strings,
a type of topological defect that arises when a U(1) gauge
symmetry is spontaneously broken\cite{vilenkin}. We can obtain the
correction to the density of states induced by a set of defects
with arbitrary opening angle by coupling the Dirac equation to a
curved space with an appropriate metric as described in ref.
\cite{AHO97}. The metric of a two dimensional space in presence of
a single cosmic string in polar coordinates is:
\begin{equation}
ds^{2}=-dt^{2}+dr^{2}+c^{2}r^{2}d\theta^{2},\label{metric}
\end{equation}
where the parameter $c$ is a constant related to the deficit angle
by $c=1-b$.

The dynamics of a massless Dirac spinor in a curved spacetime is
governed by the Dirac equation:
\begin{equation}
i\gamma^{\mu}\nabla_{\mu}\psi=0 \label{dircurv}
\end{equation}
The difference with the flat space  lies in the definition
of the $\gamma$ matrices that satisfy generalized anticommutation
relations
\begin{equation}
\{\gamma^{\mu},\gamma^{\nu}\}=2g^{\mu\nu},
\end{equation}
where $g_{\mu \nu}$ is given by (\ref{metric}),
and in the covariant derivative operator, defined as
\begin{equation}
\nabla_{\mu}=\partial_{\mu}-\Gamma_{\mu}
\end{equation}
where $\Gamma_{\mu}$ is the spin connection of the
spinor field that can be
calculated using the tetrad formalism\cite{birrell}.
\begin{figure}
  \begin{center}
    \epsfig{file=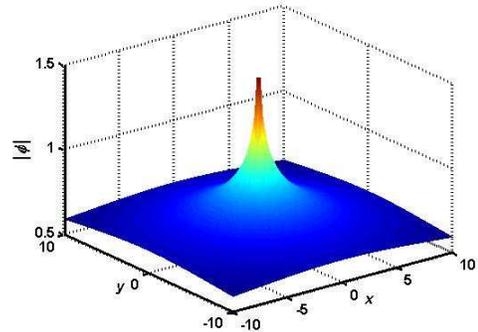,height=5cm}
    \caption{Electronic density  around a conical defect.
    }
    \label{1string}
\end{center}
\end{figure}
Fig. \ref{1string} shows the solution of the Dirac
equation (\ref{dircurv}) in the presence
of a single defect with a positive deficit angle
(positive curvature).
The electronic density is strongly peaked at the
position of the defect suggesting a bound state but
the behavior at large distances is a power law with an angle-dependent
exponent less than two
which corresponds to a non normalizable wave function.
This  behavior  is similar to the
one found in the case of a single vacancy\cite{PGLPC06}
and suggests that a
system with a number of this defects with overlapping
wave functions will be metallic.

The case of   a single cosmic string which
represents a deficit angle in the space can be generalized
to describe seven membered rings representing an angle
surplus by considering a value for the deficit angle c
greater than $1$.  This situation is non-physical from a general
relativity viewpoint as it would correspond to a string with negative mass
density but it makes perfect sense in our case. The scenario can also
be generalized to describe an arbitrary number of pentagons and
heptagons by using  the following metric:
\begin{equation}
ds^{2}=-dt^{2}+e^{-2\Lambda(x,y)}(dx^{2}+dy^{2}),\label{genmetric}
\end{equation}
where $\Lambda(\textbf{r})=\sum^{N}_{i=1}4\mu_{i}\log(r_{i})$ and
$r_{i}=[(x-a_{i})^{2}+(y-b_{i})^{2}]^{1/2}.$ This metric describes
the space-time around N parallel cosmic strings, located at the
points $(a_{i},b_{i})$. The parameters $\mu_{i}$ are related to the
angle defect or surplus by the relationship $c_{i}=1-4\mu_{i}$ in
such manner that if $c_{i}<1 (>1)$ then $\mu_{i}>0 (<0)$.

From equation (\ref{dircurv}) we can
write down
the Dirac equation for the electron propagator, $S_{F}(x,x')$:
\begin{equation}
i\gamma^{\mu}({\bf r})(\partial_{\mu}-\Gamma_{\mu})S_{F}(x,x')=
\frac{1}{\sqrt{-g}}\;\delta^{3}(x-x'),
\label{propcurv}
\end{equation}
where $x=(t, {\bf r})$.
The local density of states $N(\omega,\textbf{r})$ is obtained from (\ref{propcurv})
by Fourier transforming the time component and taking the limit
${\bf r}\to {\bf r'}$:
\begin{equation}
N(\omega,\textbf{r})=Im
Tr S_{F}(\omega,\textbf{r},\textbf{r}).\label{LDOS}
\end{equation}
We solve eq. (\ref{LDOS}) considering the  curvature induced
by the defects as a perturbation of the flat graphene layer.
The details of the calculation will be given elsewere\cite{CV06}.
Here we will show the results obtained.

\begin{figure}
  \begin{center}
    \epsfig{file=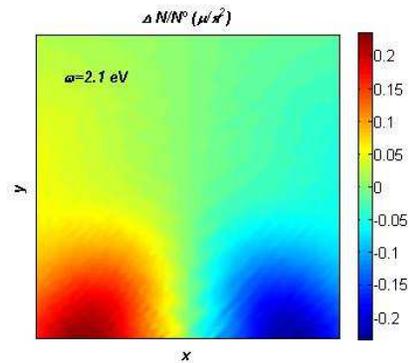,height=5cm}
    \caption{(Color online) Image of the correction to the local density of states in a large portion
    of the plane with a heptagon-pentagon  pair located out of plane at position
    $[0.3, -5], [-0.3, -5]$ .
    }
    \label{pair2}
\end{center}
\end{figure}

\begin{figure}
  \begin{center}
    \epsfig{file=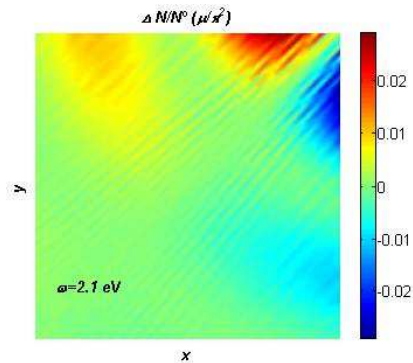,height=5cm}
    \caption{(Color online) Correction to the local density of states around a Stone-Wales defect
    located out of the image in the upper right corner.}
    \label{stone-wales2}
\end{center}
\end{figure}

{\it Results.}
Fig. \ref{pair2} shows the correction to the local density of states
at energy $w=2.1 eV$ and for a large region of the graphene plane with a
pentagon (p)-heptagon (h) pair located out of plane
at $p=[-0.3,-5], h=[0.3,-5]$. The LDoS is normalized
with the clean DoS of graphene at the given energy. The color code is indicated in
the figure: green stands for the DOS of perfect graphene at the given energy
and red (blue) indicates an accumulation (depletion) of the density in the
area. The correction obtained is of the order of a few percent.
We can see that pentagonal (heptagonal) rings  enhance
(deppress) the electron density. A similar result has been obtained in
\cite{TT94} with numerical simulations. It is to note that a somehow
contradictory result was obtained in \cite{AFM98} where they studied
the electrostatics of a graphene plane with defects. They found that
disclinations corresponding to rings with more (less) than six carbon atoms function
as attractors (repellent) to point charges. It is obvious that this
issue needs further investigation.  The dipolar character of the defect is
clear from the figure and makes this type of defects unique. The effects
produced will extend spatially much beyond the scale of the defect what
can be related with
recent Electrostatic Force Microscopy (EFM) measurements
that indicate large potential differences
between micrometer large regions on the surface of highly oriented graphite
that do not show inhomogeneities in STM\cite{GE06}.
The interference patterns  in the DOS depend on the position and on the relative
orientation of the dipoles. Fig. \ref{stone-wales2}
shows the patterns produced by a Stone-Wales defect (two pairs
of pentagons ($p_i$) heptagons $(h_i)$)
located out of plane at positions $p_1=[5,5.6], p_2=[6.5,6],
h_1=[6,5.5], h_2=[6,6.9]$  for a frequency $w=2.1 eV$.

{\it A random distribution of defects.}
A random distribution of
topological defects can be described in the continuous model
by a (non abelian) random gauge field.
The disorder is defined by a single dimensionless quantity,
$\Delta$, which is proportional to the average fluctuations
of the field:
\begin{equation}
\langle {\bf \vec{A}} ( {\bf \vec{r}} ) {\bf \vec{A}} (
{\bf \vec{r}'} ) \rangle = \Delta \delta^2 ( {\bf \vec{r}} -
{\bf \vec{r}'} )
\label{delta}
\end{equation}
The statistical properties of the gauge field induced by
topological defects were analyzed in ref. \cite{GGV01} following a
renormalization group scheme and a complete phase diagram as a
function of disorder and Coulomb interaction was obtained in
\cite{SGV05}.

In ref. \cite{GGV01} it was shown that
when  pentagons and heptagons are bound in dislocations
with average distance $b$,  the vector field of a vortex-antivortex
dipole decays as $r^{-2}$ and the behavior of $\Delta$ was found to be:
\begin{equation}
\Delta \propto \Phi_0^2 n b^2
\label{disloc}
\end{equation}
where $ n$ is the density of dislocations.

Assuming that random fields induced by topological defects have
the same statistical properties to those with gaussian disorder with
the same value of $\Delta$,
a renormalization group (RG) analysis using the replica trick
gives rise to an effective interaction between
fermion fields in different replicas. The resulting self-energy
is logarithmically divergent what  can be interpreted
as a renormalization of the density of states\cite{GGV01}.
The results obtained in this paper show
a qualitative picture of the corrected DoS. The influence of
these defects on the localization in graphene has been
analyzed recently in \cite{MG06}.

The type of disorder analyzed in  this article belongs to the
random gauge field in the classification of \cite{SGV05}. There it
was shown that the combination of this type of disorder with the
unscreened Coulomb interaction give rise to a line of infrared
stable fixed points. Topological defects give rise also to long
range correlated disorder  whose properties will be analyzed
elsewhere\cite{CV06}.

{\it Conclusions and open problems.} We have presented in this
article results on the electronic structure of a curved graphene
sheet with pentagon-heptagon pairs that we expect to be present in
the samples that are subject of present investigation. The
formalism can also apply to clean samples of graphite showing
strong anisotropy and to curved samples of graphene with an
arbitrary number of defects with any sign of the curvature. The
model can also be  used to study transport properties of the
curved graphene sheets and of nanographite and results will be
presented in \cite{CV06}. The main achievement  of this paper is
the proposal of a formalism that can describe the corrugations
observed in the experiments and predicts charge inhomogeneities
that can be used to characterize the samples. We have computed the
local density of states with the geometrical formalism  found that
it is enhanced around defects which induce positive curvature in
the lattice while the charge is "repelled" from regions with
negative curvature. Heptagon-pentagon pairs that keep the graphene
sheet flat in the long range behave as dipoles and give rise to
characteristic modulations of the DOS that can be observed by STM
and can explain recent experiments. The magnitude of the
oscillations increases with the frequency and the characteristic
patterns could be used to characterize the graphene samples. The
features predicted in this work should also be observable in other
layered materials with similar structure as boron
nitride\cite{TTM01}. The present analysis can help to clarify the
issue of the analysis and interpretation of STM images.

We thank B. Valenzuela, P. Guinea and P. Esquinazi for many illuminating discussions.
A. C. thanks N. Ozdemir for kindly  providing details of the calculations
in ref. \cite{AHO97}.
Funding from MCyT (Spain) through grant MAT2002-0495-C02-01
and from the European Union Contract No. 12881 (NEST) is
acknowledged.

\bibliography{EPL}

\newcommand{\npb}{Nucl. Phys. B}\newcommand{\adv}{Adv.
  Phys.}\newcommand{\epl}{Europhys. Lett.}
\begin{thebibliography}{31}
\expandafter\ifx\csname natexlab\endcsname\relax\def\natexlab#1{#1}\fi
\expandafter\ifx\csname bibnamefont\endcsname\relax
  \def\bibnamefont#1{#1}\fi
\expandafter\ifx\csname bibfnamefont\endcsname\relax
  \def\bibfnamefont#1{#1}\fi
\expandafter\ifx\csname citenamefont\endcsname\relax
  \def\citenamefont#1{#1}\fi
\expandafter\ifx\csname url\endcsname\relax
  \def\url#1{\texttt{#1}}\fi
\expandafter\ifx\csname urlprefix\endcsname\relax\def\urlprefix{URL }\fi
\providecommand{\bibinfo}[2]{#2}
\providecommand{\eprint}[2][]{\url{#2}}

\bibitem[{\citenamefont{Novoselov et~al.}(2005)}]{Netal05}
\bibinfo{author}{\bibfnamefont{K.~S.} \bibnamefont{Novoselov}}
  \bibnamefont{et~al.}, \bibinfo{journal}{Nature}
  \textbf{\bibinfo{volume}{438}}, \bibinfo{pages}{197} (\bibinfo{year}{2005}).

\bibitem[{\citenamefont{Zhang et~al.}(2005)}]{Zetal05}
\bibinfo{author}{\bibfnamefont{Y.}~\bibnamefont{Zhang}} \bibnamefont{et~al.},
  \bibinfo{journal}{Nature} \textbf{\bibinfo{volume}{438}},
  \bibinfo{pages}{201} (\bibinfo{year}{2005}).

\bibitem[{\citenamefont{Semenoff}(1984)}]{S84}
\bibinfo{author}{\bibfnamefont{G.~V.} \bibnamefont{Semenoff}},
  \bibinfo{journal}{Phys. Rev. Lett.} \textbf{\bibinfo{volume}{53}},
  \bibinfo{pages}{2449} (\bibinfo{year}{1984}).

\bibitem[{\citenamefont{Haldane}(1988)}]{Hal88}
\bibinfo{author}{\bibfnamefont{F.~D.~M.} \bibnamefont{Haldane}},
  \bibinfo{journal}{Phys. Rev. Lett.} \textbf{\bibinfo{volume}{61}},
  \bibinfo{pages}{2015} (\bibinfo{year}{1988}).

\bibitem[{\citenamefont{Gonz\'alez et~al.}(1996)\citenamefont{Gonz\'alez,
  Guinea, and Vozmediano}}]{GGV96}
\bibinfo{author}{\bibfnamefont{J.}~\bibnamefont{Gonz\'alez}},
  \bibinfo{author}{\bibfnamefont{F.}~\bibnamefont{Guinea}}, \bibnamefont{and}
  \bibinfo{author}{\bibfnamefont{M.~A.~H.} \bibnamefont{Vozmediano}},
  \bibinfo{journal}{Phys. Rev. Lett.} \textbf{\bibinfo{volume}{77}},
  \bibinfo{pages}{3589} (\bibinfo{year}{1996}).

\bibitem[{\citenamefont{Khveshchenko}(2001)}]{Khv01}
\bibinfo{author}{\bibfnamefont{D.~V.} \bibnamefont{Khveshchenko}},
  \bibinfo{journal}{Phys. Rev. Lett.} \textbf{\bibinfo{volume}{87}},
  \bibinfo{pages}{246802} (\bibinfo{year}{2001}).

\bibitem[{\citenamefont{Kopelevich et~al.}(2003)}]{Ketal03}
\bibinfo{author}{\bibfnamefont{Y.}~\bibnamefont{Kopelevich}}
  \bibnamefont{et~al.}, \bibinfo{journal}{\prl} \textbf{\bibinfo{volume}{90}},
  \bibinfo{pages}{156402} (\bibinfo{year}{2003}).

\bibitem[{\citenamefont{Novoselov et~al.}(2004)}]{Netal04}
\bibinfo{author}{\bibfnamefont{K.~S.} \bibnamefont{Novoselov}}
  \bibnamefont{et~al.}, \bibinfo{journal}{Science}
  \textbf{\bibinfo{volume}{306}}, \bibinfo{pages}{666} (\bibinfo{year}{2004}).

\bibitem[{\citenamefont{Esquinazi et~al.}(2003)}]{Eetal03}
\bibinfo{author}{\bibfnamefont{P.}~\bibnamefont{Esquinazi}}
  \bibnamefont{et~al.}, \bibinfo{journal}{Phys. Rev. Lett.}
  \textbf{\bibinfo{volume}{91}}, \bibinfo{pages}{227201}
  (\bibinfo{year}{2003}).

\bibitem[{\citenamefont{Ajayan et~al.}(1998)}]{Aetal98}
\bibinfo{author}{\bibfnamefont{P.~M.} \bibnamefont{Ajayan}}
  \bibnamefont{et~al.}, \bibinfo{journal}{Phys. Rev. Lett.}
  \textbf{\bibinfo{volume}{81}}, \bibinfo{pages}{1437} (\bibinfo{year}{1998}).

\bibitem[{\citenamefont{Hashimoto et~al.}(2004)}]{Hetal04}
\bibinfo{author}{\bibfnamefont{A.}~\bibnamefont{Hashimoto}}
  \bibnamefont{et~al.}, \bibinfo{journal}{Nature}
  \textbf{\bibinfo{volume}{430}}, \bibinfo{pages}{870} (\bibinfo{year}{2004}).

\bibitem[{\citenamefont{Lu et~al.}(2006)}]{GE06}
\bibinfo{author}{\bibfnamefont{Y.}~\bibnamefont{Lu}} \bibnamefont{et~al.},
  \bibinfo{journal}{Phys. Rev. Lett.} \textbf{\bibinfo{volume}{97}},
  \bibinfo{pages}{076805} (\bibinfo{year}{2006}).

\bibitem[{\citenamefont{Morozov et~al.}(2006)}]{Metal06}
\bibinfo{author}{\bibfnamefont{S.}~\bibnamefont{Morozov}} \bibnamefont{et~al.},
  \bibinfo{journal}{Phys. Rev. Lett.} \textbf{\bibinfo{volume}{97}},
  \bibinfo{pages}{016801} (\bibinfo{year}{2006}).

\bibitem[{\citenamefont{Wallace}(1947)}]{W47}
\bibinfo{author}{\bibfnamefont{P.~R.} \bibnamefont{Wallace}},
  \bibinfo{journal}{Phys. Rev.} \textbf{\bibinfo{volume}{71}},
  \bibinfo{pages}{622} (\bibinfo{year}{1947}).

\bibitem[{\citenamefont{Slonczewski and Weiss}(1958)}]{SW58}
\bibinfo{author}{\bibfnamefont{J.~C.} \bibnamefont{Slonczewski}}
  \bibnamefont{and} \bibinfo{author}{\bibfnamefont{P.~R.} \bibnamefont{Weiss}},
  \bibinfo{journal}{Phys. Rev.} \textbf{\bibinfo{volume}{109}},
  \bibinfo{pages}{272} (\bibinfo{year}{1958}).

\bibitem[{\citenamefont{Gonz\'alez et~al.}(1992)\citenamefont{Gonz\'alez,
  Guinea, and Vozmediano}}]{GGV92}
\bibinfo{author}{\bibfnamefont{J.}~\bibnamefont{Gonz\'alez}},
  \bibinfo{author}{\bibfnamefont{F.}~\bibnamefont{Guinea}}, \bibnamefont{and}
  \bibinfo{author}{\bibfnamefont{M.~A.~H.} \bibnamefont{Vozmediano}},
  \bibinfo{journal}{Phys. Rev. Lett.} \textbf{\bibinfo{volume}{69}},
  \bibinfo{pages}{172} (\bibinfo{year}{1992}).

\bibitem[{\citenamefont{Gonz\'alez et~al.}(1993)\citenamefont{Gonz\'alez,
  Guinea, and Vozmediano}}]{GGV93}
\bibinfo{author}{\bibfnamefont{J.}~\bibnamefont{Gonz\'alez}},
  \bibinfo{author}{\bibfnamefont{F.}~\bibnamefont{Guinea}}, \bibnamefont{and}
  \bibinfo{author}{\bibfnamefont{M.~A.~H.} \bibnamefont{Vozmediano}},
  \bibinfo{journal}{Nucl. Phys. B} \textbf{\bibinfo{volume}{406 [FS]}},
  \bibinfo{pages}{771} (\bibinfo{year}{1993}).

\bibitem[{\citenamefont{Gonz\'alez et~al.}(1994)\citenamefont{Gonz\'alez,
  Guinea, and Vozmediano}}]{GGV94}
\bibinfo{author}{\bibfnamefont{J.}~\bibnamefont{Gonz\'alez}},
  \bibinfo{author}{\bibfnamefont{F.}~\bibnamefont{Guinea}}, \bibnamefont{and}
  \bibinfo{author}{\bibfnamefont{M.~A.~H.} \bibnamefont{Vozmediano}},
  \bibinfo{journal}{Nucl. Phys. B} \textbf{\bibinfo{volume}{424 [FS]}},
  \bibinfo{pages}{595} (\bibinfo{year}{1994}).

\bibitem[{\citenamefont{Gonz\'alez et~al.}(2001)\citenamefont{Gonz\'alez,
  Guinea, and Vozmediano}}]{GGV01}
\bibinfo{author}{\bibfnamefont{J.}~\bibnamefont{Gonz\'alez}},
  \bibinfo{author}{\bibfnamefont{F.}~\bibnamefont{Guinea}}, \bibnamefont{and}
  \bibinfo{author}{\bibfnamefont{M.~A.~H.} \bibnamefont{Vozmediano}},
  \bibinfo{journal}{Phys. Rev. B} \textbf{\bibinfo{volume}{63}},
  \bibinfo{pages}{134421} (\bibinfo{year}{2001}).

\bibitem[{\citenamefont{Cortijo and Vozmediano}(2007)}]{CV06}
\bibinfo{author}{\bibfnamefont{A.}~\bibnamefont{Cortijo}} \bibnamefont{and}
  \bibinfo{author}{\bibfnamefont{M.~A.~H.} \bibnamefont{Vozmediano}},
  \bibinfo{journal}{Nucl. Phys. B} \textbf{\bibinfo{volume}{763}},
  \bibinfo{pages}{293} (\bibinfo{year}{2007}).

\bibitem[{\citenamefont{Pudlak et~al.}(2006)\citenamefont{Pudlak, Pincak, and
  Osipov}}]{PPO06}
\bibinfo{author}{\bibfnamefont{M.}~\bibnamefont{Pudlak}},
  \bibinfo{author}{\bibfnamefont{R.}~\bibnamefont{Pincak}}, \bibnamefont{and}
  \bibinfo{author}{\bibfnamefont{V.}~\bibnamefont{Osipov}}
  (\bibinfo{year}{2006}), \eprint{cond-mat/0602520}.

\bibitem[{\citenamefont{Furtado et~al.}(2006)\citenamefont{Furtado, Moraes, and
  de~M.~Carvalho}}]{FMC06}
\bibinfo{author}{\bibfnamefont{C.}~\bibnamefont{Furtado}},
  \bibinfo{author}{\bibfnamefont{F.}~\bibnamefont{Moraes}}, \bibnamefont{and}
  \bibinfo{author}{\bibfnamefont{A.~M.} \bibnamefont{de~M.~Carvalho}}
  (\bibinfo{year}{2006}), \eprint{cond-mat/0601391}.

\bibitem[{\citenamefont{Vilenkin and Shellard}(2000)}]{vilenkin}
\bibinfo{author}{\bibfnamefont{A.}~\bibnamefont{Vilenkin}} \bibnamefont{and}
  \bibinfo{author}{\bibfnamefont{E.~P.~S.} \bibnamefont{Shellard}},
  \emph{\bibinfo{title}{Cosmic Strings and Other Topological Defects}}
  (\bibinfo{publisher}{Cambridge University Press}, \bibinfo{year}{2000}).

\bibitem[{\citenamefont{Aliev et~al.}(1997)\citenamefont{Aliev, H\"{O}rtacsu,
  and Ozdemir}}]{AHO97}
\bibinfo{author}{\bibfnamefont{A.~N.} \bibnamefont{Aliev}},
  \bibinfo{author}{\bibfnamefont{M.}~\bibnamefont{H\"{O}rtacsu}},
  \bibnamefont{and} \bibinfo{author}{\bibfnamefont{N.}~\bibnamefont{Ozdemir}},
  \bibinfo{journal}{Class. Quantum. Grav.} \textbf{\bibinfo{volume}{14}},
  \bibinfo{pages}{3215} (\bibinfo{year}{1997}).

\bibitem[{\citenamefont{Birrell and Davis}(1982)}]{birrell}
\bibinfo{author}{\bibnamefont{Birrell}} \bibnamefont{and}
  \bibinfo{author}{\bibnamefont{Davis}}, \emph{\bibinfo{title}{Quantum fields
  in curved space}} (\bibinfo{publisher}{Cambridge University Press},
  \bibinfo{year}{1982}).

\bibitem[{\citenamefont{Pereira et~al.}(2006)}]{PGLPC06}
\bibinfo{author}{\bibfnamefont{V.~M.} \bibnamefont{Pereira}}
  \bibnamefont{et~al.}, \bibinfo{journal}{Phys. Rev. Lett.}
  \textbf{\bibinfo{volume}{96}}, \bibinfo{pages}{036801}
  (\bibinfo{year}{2006}).

\bibitem[{\citenamefont{Tamura and Tsukada}(1994)}]{TT94}
\bibinfo{author}{\bibfnamefont{R.}~\bibnamefont{Tamura}} \bibnamefont{and}
  \bibinfo{author}{\bibfnamefont{M.}~\bibnamefont{Tsukada}},
  \bibinfo{journal}{Phys. Rev. B.} \textbf{\bibinfo{volume}{49}},
  \bibinfo{pages}{7697} (\bibinfo{year}{1994}).

\bibitem[{\citenamefont{Azevedo et~al.}(1998)\citenamefont{Azevedo, Furtado,
  and Moraes}}]{AFM98}
\bibinfo{author}{\bibfnamefont{S.}~\bibnamefont{Azevedo}},
  \bibinfo{author}{\bibfnamefont{C.}~\bibnamefont{Furtado}}, \bibnamefont{and}
  \bibinfo{author}{\bibfnamefont{F.}~\bibnamefont{Moraes}},
  \bibinfo{journal}{Physica Status Solidi b} \textbf{\bibinfo{volume}{207}},
  \bibinfo{pages}{387} (\bibinfo{year}{1998}).

\bibitem[{\citenamefont{Stauber et~al.}(2005)\citenamefont{Stauber, Guinea, and
  Vozmediano}}]{SGV05}
\bibinfo{author}{\bibfnamefont{T.}~\bibnamefont{Stauber}},
  \bibinfo{author}{\bibfnamefont{F.}~\bibnamefont{Guinea}}, \bibnamefont{and}
  \bibinfo{author}{\bibfnamefont{M.~A.~H.} \bibnamefont{Vozmediano}},
  \bibinfo{journal}{Phys. Rev. B} \textbf{\bibinfo{volume}{71}},
  \bibinfo{pages}{041406(R)} (\bibinfo{year}{2005}).

\bibitem[{\citenamefont{Morpurgo and Guinea}(2006)}]{MG06}
\bibinfo{author}{\bibfnamefont{A.~F.} \bibnamefont{Morpurgo}} \bibnamefont{and}
  \bibinfo{author}{\bibfnamefont{F.}~\bibnamefont{Guinea}},
  \bibinfo{journal}{Phys. Rev. Lett.} \textbf{\bibinfo{volume}{97}},
  \bibinfo{pages}{196804} (\bibinfo{year}{2006}).

\bibitem[{\citenamefont{Terrones et~al.}(2001)\citenamefont{Terrones, Terrones,
  and Mor\'an-L\'opez}}]{TTM01}
\bibinfo{author}{\bibfnamefont{H.}~\bibnamefont{Terrones}},
  \bibinfo{author}{\bibfnamefont{M.}~\bibnamefont{Terrones}}, \bibnamefont{and}
  \bibinfo{author}{\bibfnamefont{J.~L.} \bibnamefont{Mor\'an-L\'opez}},
  \bibinfo{journal}{Current Science} \textbf{\bibinfo{volume}{81}},
  \bibinfo{pages}{1011} (\bibinfo{year}{2001}).

\end{thebibliography}
\end{document}